\documentclass[11pt]{article}

\usepackage{palatino}

\RequirePackage[OT1]{fontenc} 

\usepackage{parskip}

\usepackage[slantedGreek]{mathpazo}

\usepackage{amsfonts}
\usepackage{setspace}
\usepackage{amssymb}
\usepackage{amsthm}
\usepackage{graphicx}
\usepackage{amsmath}
\usepackage{mathtools}
\usepackage{lscape}
\usepackage{subcaption}
\usepackage{suffix}
\usepackage{color}

\usepackage{booktabs}
\usepackage{longtable}
\usepackage{array}
\usepackage{multirow}
\usepackage[table]{xcolor}
\usepackage{wrapfig}
\usepackage{float}
\floatstyle{plaintop}
\restylefloat{table}
\usepackage{colortbl}
\usepackage{pdflscape}
\usepackage{tabu}
\usepackage{threeparttable}
\usepackage{threeparttablex}
\usepackage[normalem]{ulem}
\usepackage{makecell}
\usepackage{microtype}
\usepackage{float} 
\usepackage[margin=1in,footskip=.25in]{geometry} 
\RequirePackage[colorlinks,citecolor=black,urlcolor=black, linkcolor = black]{hyperref} 

\usepackage{color}
\definecolor{DarkBlue}{rgb}{0.1,0.1,0.5}
\definecolor{Red}{rgb}{0.9,0.0,0.1}
\definecolor{Navy}{rgb}{0.00,0.00,0.30}
\definecolor{Yellow}{rgb}{1.00,1.00,0.00}
\definecolor{Gold}{rgb}{1.00,0.84,0.00}
\definecolor{Lightgoldenrod}{rgb}{0.93,0.87,0.51}
\definecolor{Goldenrod}{rgb}{0.85,0.65,0.13}
\definecolor{Black2}{rgb}{0.00,0.00,0.00}
\definecolor{orange}{rgb}{0.85,0.65,0.13}
\definecolor{SkyBlue}{rgb}{0.941176,0.972549,1.}
\definecolor{MyLightMagenta}{cmyk}{0.1,0.8,0,0.1}
\usepackage{xcolor}

\usepackage{graphicx}

\usepackage[authoryear]{natbib}

\usepackage{titlesec}

\allowdisplaybreaks

\renewcommand{\u}{{\bf u}}

\newcommand{\W}{{\bf W}}
\newcommand{\X}{{\bf X}}
\newcommand{\Y}{{\bf Y}}

\newcommand{\ben}{\begin{enumerate}}
\newcommand{\een}{\end{enumerate}}
\newcommand{\beq}{\begin{equation}}
\newcommand{\eeq}{\end{equation}}
\newcommand{\bde}{\begin{description}}
\newcommand{\ede}{\end{description}}

\newcommand{\abs}[1]{\lvert#1\rvert}

\newtheoremstyle{slplain}
  {1\baselineskip\@plus.2\baselineskip\@minus.2\baselineskip}
  {.5\baselineskip\@plus.2\baselineskip\@minus.2\baselineskip}
  {\slshape}
  {}
  {\bfseries}
  {.}
  { }
  {}

\usepackage{booktabs,array}

\newcount\rowc

\graphicspath{{../../figures/}{../figures/}{./figures/}{./}}

\numberwithin{equation}{section}

\usepackage[inline]{enumitem}
\setlist*[enumerate]{label=(\roman*)}

\theoremstyle{plain}

\theoremstyle{definition}
\newtheorem{definition}{Definition}[section]

\theoremstyle{remark}

\title{Quantifying the Effect of Socio-Economic Predictors and Built Environment on Mental Health Events in Little Rock, AR \thanks{This study is a continuation of work presented in the first author's MS thesis and is a work-in-progress draft.}}

\author{Alfieri Ek \\
	Department of Mathematical Sciences, University of Arkansas - Fayetteville \\
	and \\
	Samantha Robinson \\
	Department of Mathematical Sciences, University of Arkansas - Fayetteville \\
	and \\
    Grant Drawve \\
	Department of Sociology and Criminology, University of Arkansas - Fayetteville \\
	and \\
	Jyotishka Datta \footnote{Corresponding author} \\
    Department of Statistics, Virginia Polytechnic Institute and State University}

\begin{document}

\maketitle

\begin{abstract}
Proper allocation of law enforcement resources remains a critical issue in crime prediction and prevention that operates by characterizing  spatially aggregated  crime activities and a multitude of predictor variables of interest. Despite the critical nature of proper resource allocation for mental health incidents, there has been little progress in statistical modeling of the geo-spatial nature of mental health events in Little Rock, Arkansas. In this article, we provide insights into the spatial nature of mental health data from Little Rock, Arkansas between 2015 and 2018, under a supervised spatial modeling framework while extending the popular risk terrain modeling \citep{caplan2011risk, caplan2015risk, drawve2016metric} approach. We provide evidence of spatial clustering and identify the important features influencing such heterogeneity via a spatially informed hierarchy of generalized linear models, spatial regression models and a tree based method, \textit{viz.}, Poisson regression, spatial Durbin error model, Manski model and Random Forest. The insights obtained from these different models are presented here along with their relative predictive performances. The inferential tools developed here can be used in a broad variety of spatial modeling contexts and have the potential to aid both law enforcement agencies and the city in properly allocating resources.
\end{abstract}

\section{Introduction}\label{chap:intro}
Over the last two decades, law enforcement agencies are relying more and more on statistical tools to build an objective criminal justice system, leading to a meteoric rise of ``predictive policing", loosely defined as ``\textit{the application of analytical techniques - particularly quantitative techniques - to identify likely targets for police intervention and prevent crime or solve past crimes by making statistical predictions}" \citep{perry2013predictive}. The proposed algorithms and methods attempt to uncover and exploit different aspects of crime activities data. For example, \citet{gotway1997generalized} use a spatial generalized linear model, that has been extended both by considering the temporal pattern as well as a non-linear modeling approach using generalized additive modeling in ST-GAM or LST-GAM \citep{wang2012spatio}. In a series of papers, \citet{mohler2011self, mohler2013modeling, mohler2015randomized} propose a self-exciting point process model that treats near-repeat nature of crimes \citep{townsley2000repeat} as aftershocks of an earthquake. This is the main driving force behind the popular crime forecasting software called PredPol (\url{https://predpol.com/}) that has been since adopted by many policing agencies throughout the US. 

Apart from increasing the accuracy of prediction of future crime, it is also important to understand which geographical factors significantly contribute to crime. Such knowledge can inform a plan for allocating resources or making policy changes to either counteract the effect of a `risky' place or increase the intensity or presence of a `protective' place. This is also closely related to the goal of ensuring that a prediction rule does not suffer from algorithmic or systemic biases. This is particularly important, as with the increase in complexity and use of such data-based tools, there is growing concern and additional effort devoted to reducing the racial disparities in predictive policing, while producing dynamic and real-time forecasts and insights about spatio-temporal crime activities. For example, using a combination of demographically representative synthetic data and survey data on drug use, \citet{lum2016predict} point out that predictive policing estimates based on biased policing records often accentuate the racial bias instead of removing it. A natural solution seems to be the risk terrain modeling (RTM) framework of \citet{caplan2011risk}, that uses a simple but interpretable approach. In RTM, a separate map layer is created for each predictor, that are then combined to produce a composite map where contribution or importance of each factor can be evaluated in a model-based way. 

We start with a brief review of the existing statistical methodology behind the most common crime forecasting tools. 

\subsection{Literature Review}

\noindent \textbf{Self-exciting Point Process:} One of the popular statistical approaches to modeling criminal activities is self-exciting processes \citep{mohler2011self, mohler2013modeling, mohler2015randomized} that is characterized by the increasing probability of repeated events following an event, similar to aftershocks of an earthquake. Here the intensity of a discrete-time point process (criminal activities, in this context) is determined as a log-Gaussian Cox process (LGCP) whose intensity is self-excited by occurrence of many events in a short time-window. \citet{mohler2015randomized} found their approach outperformed a dedicated crime analyst who relied on existing intel and hotspot mapping.

\noindent \textbf{Generalized Additive Modeling for Spatio-temporal Data:} \cite{wang2012spatio} developed a more sophisticated model using a generalized additive modeling for spatio-temporal data (ST-GAM) that can be thought of as an extension of grid-based regression approaches that can account for non-linear relationships. Here, spatio-temporal features include previous crime activities, socio-economic and built-environment features at the grid-cell resolution indexed over time, and \cite{wang2012spatio} showed that their method outperforms spatial Generalized Linear Model (GLM) \citep{gotway1997generalized} where temporal information is not incorporated. 

\noindent \textbf{Risk Terrain Modeling:} Risk terrain modeling, henceforth abbreviated as RTM, \citep{caplan2011risk, caplan2015risk, drawve2016metric} is a class of statistical methods that combines geographic features such as built, physical-environment and socioeconomic variables in a supervised learning set-up to provide insights and forecasts for crime activities at a chosen grid-level based on the proximity to features and social factors or density of features. A typical RTM approach involves three steps: (1) identify potentially relevant factors for the spatial varying response variable, (2) assign a value for each factor considered for each location or grid-cell spanning a common geography, and (3) combine the factor-specific raster maps in a supervised regression framework so that each factor can be judged in terms of its relevance for the crime outcome. The RTM approach, like several other models, alleviates some racial disparity concerns by moving the focus of the modeling approach from people to places. However, there are some key advantages of the RTM approach over the LST-GAM or Hawkes process based algorithms. Firstly, the underlying statistical methodology for RTM immediately provides interpretability to the factors influencing spatial clustering of crime or other response variables. Secondly, the raster-map based modeling framework lets us easily incorporate different machine learning and statistical tools of choice depending on their performance for a given jurisdiction. In this paper, we use Poisson GLM, spatial error model and random forest, but it is straightforward to add any number of methodologies to the mix and choose the best performing method or combine the disparate tools in an ensemble learning framework. 

While these developments have been mostly focused on crime prediction and prevention, there is relatively less emphasis on other spatial events such as mental health calls that also require resource allocation from law enforcement agencies or the city. The goal of this paper is to extend the powerful and interpretable statistics and machine learning methodologies under the general umbrella of risk terrain modeling to the geo-spatial predictive modeling of mental health call locations in Little Rock, AR.

The outline of the paper is as follows: in Section \ref{chap:sp}, we describe the modeling approach and the different methodologies used in developing the risk terrain model for mental health calls. In Section \ref{chap:mental} we illustrate the spatial clustering and other descriptive features of the data as well as demonstrating the performance of the proposed framework. Finally, in Section \ref{chap:end}, we provide some new directions for research in this area. 

\section{Spatial Forecasting}\label{chap:sp}
\subsection{Modeling Approach}

Our spatial modeling and forecasting framework is similar to RTM, with a key difference being the underlying statistical methodologies. In this paper, we use the following methodologies and compare both the important predictors chosen by the model as well as their predictive performance for forecasting mental health incidents in Little Rock, Arkansas. Little Rock regularly has above average violent and property crime rates when compared to other large U.S. cities \citep{chillar2020unpacking}. Data were obtained from several city departments, including the Little Rock Police Department, through an ongoing data-sharing Memorandum of Understanding (MOU) between researchers and Little Rock. Social data were obtained from the American Community Survey (5-year estimates). Mental health incidents from 2015 through 2017 are used to predict 2018 incidents. 

\begin{description}
\item [Poisson Generalized Linear Model] The Poisson regression model belongs to a family of regression models called the generalized linear model (GLM). As a special case of the GLM family, the fitted Poisson regression model uses $\eta_i = \ln(\lambda)$ as canonical link and is of the form:
\[
\hat{y_i} =  g^{-1}(x^T_i \hat{\beta}) =  {\rm e}^{x^T_i \hat{\beta}}.
\]     
Among several link functions commonly used with the Poisson distribution, the log link function ensures that $\lambda_i \geq 0$ which is crucial for the expected value of a count outcome of the response variable (mental health incidents) \citep{Montgomery}. In terms of model interpretation, parameters may be interpreted in a probabilistic sense which arises as an advantage from the fact that Poisson regression belongs to the GLM family. Consequently, significant factors present in the fitted model may be explained in strict probabilistic terms with respective levels of uncertainty. 

\item [Random Forest] Random forest \citep{breiman2001random} falls into the non-linear/non-parametric category of supervised learning approaches known as decision trees. Decision trees are particularly known due to their inherent ease of use and interpretability in both regression and classification problems. For regression problems, which we focus on here, decision trees divide the predictor space into $J$ distinct and non-overlapping regions, $R_1,R_2,...,R_J$ also known as terminal nodes or leaves using the training data through a recursive binary splitting procedure. Note that a threshold is implemented so that the recursive binary splitting procedure ends when the number of observations at any terminal node falls below the set threshold. In addition to the preceding criteria, the aim is to obtain terminal nodes that minimize the residual sum of squares:
$$\sum^J _{j=1} \sum_{I \in R_j}(y_i - \hat{y}_{R_j})^2.$$

The results obtained are likely to over-fit the data due to the complexity of the resulting tree so, a cost-complexity pruning procedure is implemented to find a sub tree which minimizes the objective function:
 $$\sum^{|T|}_{j = 1} \sum_{i:x_i \in R_j} (y_i - \hat{y}_{R_j})^2 + \alpha |T|,$$

thereby reducing the variance at the cost of little bias for better interpretation. As a preventative measure to not over-fit the training data and control the length of the tree, the penalty factor $\alpha$ is added to $|T|$, the number of terminal nodes. The predicted response for any observation that falls into the $R^{th}_i$ region is the mean response of all observations from the training data set that are in that same terminal node. 

Single decision trees however are not as competitive when compared to other forms of linear or non-linear supervised learning models. One solution to build a more robust decision tree is known as random forests. Random forests build $B$-many trees to improve its performance using bootstrapped samples from the training data in a strategic manner that decorrelates the $B$-many trees, with the final prediction done by averaging the predictions from each of the individual trees. In the process of building each decision tree, at every stage or split, a random sample of size $m = \sqrt{p}$ predictors are chosen as candidates from the pool of $p$ predictors. As a result, strong predictors do not influence the building order of every tree (making them not look alike). This process decorrelates the $B$-many trees, as on average $\frac{p-m}{p}$ of the splits would not have such strong predictors thus reducing the variance and improving results. We refer the reader to \citet{James2014Introduction} for an in-depth discussion of random forests. In relation to crime, \citet{wheeler2021mapping} found their random forests model outperformed RTM and Kernel Density Estimations (KDE) for robbery prediction in Dallas, Texas.

\item [Spatial Econometric Model: Spatial Durbin Model] 

Data containing a location/geographic component contain spatial dependencies among observations which may lead to spatial relationships. Spatial relationships occur not only in the dependent variables (response variables), but also in the independent variables (covariates) and residual terms ($\epsilon$). The proper terms defining spatial relationships among dependent variables, independent variables and residual terms are known as endogenous interaction, exogenous interaction and error interaction respectively. A model that accounts for all spatial relationships is the \textbf{Manski model \footnote{The Manski model is also known as the Generalized Nesting Spatial Model(GNS) \citep{Elhorst2014Spatial}}}, with the form:
\begin{equation}
\Y = \delta\W\Y + \X\beta + \W\X\theta + \u; \hspace{0.5cm} \u = \lambda\W\u + \epsilon.
\label{eq:Manski}
\end{equation}
Here $\delta$ is known as the spatial autoregressive coefficient, $\lambda$ is the spatial autocorrelation coefficient, $\W$ represents the spatial weights matrix that describes the spatial configuration of the unit samples, $\X$ is a matrix of exogenous variables or covariates and lastly $\theta \text{ and } \beta$ are unknown parameters to be estimated that explain the contribution of each predictor and their spatially lagged version \citep{Elhorst2014Spatial}. 

For the purpose of this paper, both Manski and spatial Durbin error models were fitted onto the mental health spatial data. The Manksi model otherwise known as the general nesting spatial model for spatial events (mental health incidents) as a function of endogenous interactions (neighboring values or spatial lags), exogenous interactions (build environment, social factors etc.) and error interactions (spatial autocorrelation \& spatial heterogeneity). The spatial Durbin error model is a special case of a Manski model with $\delta = 0$, thus having the endogenous interactions removed. The spatial Durbin error model is of the form:
\begin{equation}
\Y = \X\beta + \W\X\theta + \u ; \hspace{0.5cm} \u = \lambda\W\u + \epsilon.
\label{eq:SDEM}
\end{equation}

\end{description}

\section{Analyzing mental health incidents in Little Rock}\label{chap:mental}

\subsection{Descriptive Statistics}
\subsubsection{Evidence of Clustering: Moran's I}

\begin{figure}[!ht]
\centering
\includegraphics[scale = 0.45]{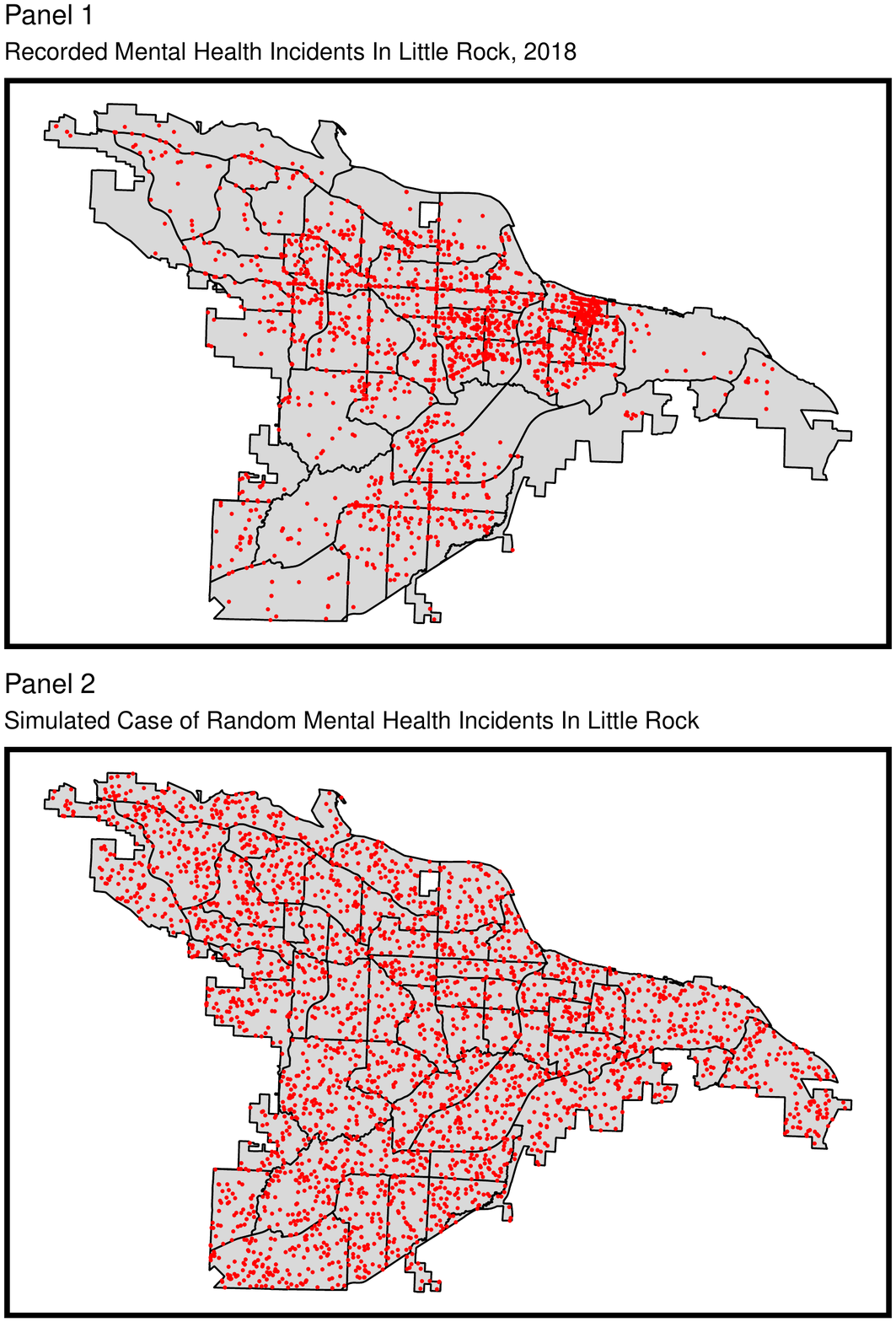}
\caption{Panel 1 shows the observed mental health incidents in Little Rock in 2018. Panel 2 shows the distribution of simulated mental health incidents following a Uniform distribution, keeping the total number of incidents fixed.}
\label{fig:Incidents}
\end{figure}

The underlying assumption at the start of this study was that mental health incident events in Little Rock were distributed as spatially heterogeneous points (\textit{i.e.,} clusters) rather than uniformly over the geographic region. To put matters into visual perspective, see Fig. \ref{fig:Incidents} where panel 1 represents the geographic locations of the recorded 2018 mental health incidents in Little Rock and panel 2 represents the same number of incidents but simulated as if they were following an uniform spatial distribution. Fig. \ref{fig:Incidents} shows the presence of spatial clusters of mental health incidents in Little Rock when compared with the uniform distribution. However, as visual comparisons could be interpreted as being subjective, we consider a measure of spatial auto-correlation to test the spatial heterogeneity. To be precise, we want to test the null hypothesis that the mental health incidents are uniformly distributed across the area of study (Little Rock) against the alternative hypothesis that they are more clustered than might be expected from usual randomness. 

Clustering, when referring to the whole spatial pattern, can be described by a global statistic for spatial auto-correlation. However, to properly identify the location of clustered and non-clustered regions, a Local Indicator of Spatial Association (LISA) must be implemented. A LISA is any statistic that provides the extent of significant spatial clustering of similar values around a given observation (\textit{i.e.,} Local Spatial Statistic). It also establishes the connection between the local and global statistic for spatial association having the sum of all local spatial statistics be proportional to the global statistic thereby allowing the decomposition of global indicators \citep{LISA}. 

Among a handful number of global tests for spatial auto-correlation including Geary's $C$ and the global Getis-$G$, Moran's $I$ is perhaps the most common global test, and is implemented in almost all common spatial toolboxes for testing auto-correlation \citep{bivand2008applied}. Spatial auto-correlation quantifies the degree to which similar features cluster and identifies their location. In the presence of spatial auto-correlation, we can predict the values of observation $i$ from the values observed at $j \in N_{i}$, the set of its proximate neighbors \citep{Pebesma2019Spatial}. As in typical correlation, Moran's $I$ value generally ranges from $-1$ to $+1$ inclusively as a result of having a normalizing factor, $n/(\sum^{n}_{i = 1} \sum^{n}_{j = 1} w_{ij})$ \citep{Moran_Range}. The contrast between spatial auto-correlation Moran's $I$ and Pearson or Spearman's correlation lie in the presence of the spatial weights matrix in Moran's $I$ statistic. The inclusion of the spatial weights matrix in Moran's $I$ enables the possibility of obtaining extreme values greater than the usual $[-1,1]$ bounds depending on the structure and composition of the weights matrix. Extreme values are obtained via the relation between the minimum and maximum eigenvalues from the spatial weights matrix. For a thorough discussion regarding the range and extreme values of Moran's $I$ we refer readers to \citep{Extreme_Morans}. A negative and significant Moran's $I$ value represents negative spatial auto-correlation indicating dissimilar values are next to each other. A positive and significant  Moran's $I$ value represents positive spatial auto-correlation indicating evidence of clustering of like values. 
 
In order to apply the spatial auto-correlation tests (both Global and Local Moran's $I$) onto the spatial data and induce a supervised learning framework, two critical prerequisite steps had to be executed, \textit{viz.} (a) identification of the $k$ nearest neighbors, and (b) assigning their respective weights using the package \textbf{spdep} \citep{SPDEP}. We first create a fishnet of grid cell size of $1000$m by $1000$m from Little Rock containing all the necessary attributes for the analysis, with each cell mapped to a centroid, which is necessary in order to extend the neighborhood criteria from contiguity to distance-based neighbors ($k$-nearest neighbors) \citep{Pebesma2019Spatial}.   

Using $k$-nearest neighbors typically leads to asymmetric neighbors. However, this is not the case here as all centroids are uniformly spaced. A key advantage of using distance-based neighbors to ordinary polygon contiguity is that it ensures that all fishnet grid cells polygon representation (centroids) have $k$ neighbors. It is common practice to use $k = 8$ or $k = 4$ neighbors which are formally know as ``Queen case" and ``Rook case". For this paper, $k = 8$ nearest neighbors were used and located using the function \emph{knearneigh} and  \emph{Knn2nb} from the package \textbf{spdep}. Following the identification of $8$-nearest neighbors for each centroid, their respective weights were assigned using the function \emph{nb2listw} from the package \textbf{spdep}. 

After the identification of the neighbors of \emph{Grid 1}, spatial weights are assigned to the list of neighbors. The entries in the weight matrix specify how much value we want to attribute to each neighbor. In this current work, we assign equal weights to each grid's neighbors, implying that each neighbor will have a corresponding weight of $\frac{1}{8}$. This weight is then used to compute the mean neighbor values as $\text{\it weight} = \frac{1}{8} \sum_{i=2}^{9} \text{\it weight for neighbor}_i$. This is equivalent to averaging over all mental health incident cases occurring within the eight neighbor grid cells. Having obtained both neighbors and their respective weights, we test for the presence of spatial auto-correlation using both Global Moran's I and Local Moran's I, as described below. 

\noindent \textbf{Global Moran's I} The process for calculating the global test for spatial auto-correlation uses local relationships between the observed spatial entity value and its defining neighbors \citep{bivand2008applied}. 

\begin{definition}[Global Moran's I] Let $y_i$ be the $i^{th}$ observation, with the mean being $\bar{y}$, and let and $w_{ij}$ be the spatial weight of the link between $i$ and $j$, then Global Moran's $I$ statistic is given by the following formula:
\begin{align*}
    I =  \frac{n}{\sum^{n}_{i = 1} \sum^{n}_{j = 1} w_{ij}} \frac{\sum^{n}_{i = 1} \sum^{n}_{j = 1} w_{ij} (y_i - \bar{y}) (y_j - \bar{y})}{ \sum^{n}_{i = 1}(y_i - \bar{y})},
\end{align*}

where $I$ represents the ratio of the product of the variable of interest, adjusted for the spatial weights used.
\label{Global}
\end{definition}

Centering on the mean is equivalent to asserting that the correct model has a constant mean, and that any remaining patterning after centering is caused by the spatial relationships encoded in the spatial weights. 

\noindent \textbf{Local Moran's $I$} Localized tests are built by breaking global measures into components which aids in the detection of clusters and hot-spots, where clusters are defined as groups of observations where neighbors have similar features and hot-spots are groups of observations with distinct neighbors \citep{bivand2008applied}. 

\begin{definition}[Local Moran's I]
Local Moran's $I_i$ values consist of the $n$ individual components added to produce the global Moran's $I$ (definition \ref{Global}): where the assumption  is that the global mean $\bar{y}$ is an accurate summary of the variable of interest $y$. Note that here we do not center the two components in the numerator, $(y_i  - \bar{y})$ and $\sum^{n}_{j = 1} w_{ij} (y_j - \bar{y})$.
\begin{align*}
    I_i = \frac{(y_i - \bar{y}) \sum^{n}_{j = 1} w_{ij} (y_j - \bar{y})}{ \frac{\sum^{n}_{i = 1} (y_i - \bar{y})^2}{n} }.
\end{align*}

\end{definition}

The global Moran's I value for the mental health incidents data was obtained as $0.22923$, computed using the function \emph{moran.test} from the \textbf{spdep} \textsc{R} package. 

To test for the significance of Global Moran's $I$ statistic, a permutation bootstrap test with $999$ simulations was conducted via the \texttt{moran.mc} function from the \texttt{spdep} \textsc{R} package. The permutation test produces a sampling distribution of the test statistic Moran's $I$ under the null hypothesis of no spatial auto-correlation, which was used to derive a (pseudo) permutation p-value, calculated using the formula: $\mathrm{p-value} = \frac{N_{\mathrm{extreme}} + 1 }{N + 1}$, where $N_{\mathrm{extreme}}$ represents the number of simulated Moran's $I$ values more extreme than the observed Moran's $I$ statistic and $N$ denotes the total number of simulations \citep{gimond2019}.

The observed value of the Global Moran's $I$ statistic produces a pseudo p-value of $1/1000 = 0.001$ when compared to the simulated values obtained from the permutation test, indicating the probability of observing a test statistic that is as or more extreme compared to the current observed Moran's $I$ value is $0.001$ under the null hypothesis $H_0$.  With the statistical significance of the Global Moran's $I$ established, a localized Moran's test was conducted to identify the location(s) of the possible mental health incident clustering using the function \textit{localmoran} from the \textbf{spdep} package. Similar to the global Moran's $I$ described above, the local Moran's $I$ evaluates the level of spatial auto-correlation among the $k$-nearest fishnet grid cells ($k = 8$, here) surrounding a given fishnet grid cell. Local Moran's test also computes the (pseudo) p-value indicating the significance of the spatial auto-correlation at the level of each fishnet grid cell. Note that na\"ively using a significance threshold of $\alpha = 0.05$ to determine which grid cells indicate a significant level of clustering will be flawed as one needs to adjust for multiple comparisons \citep{LISA}. To address the multiplicity issues, a Bonferroni adjustment was applied using the function \textit{p.adjustSP} from the \textbf{spdep} package. For the following figure, Panel 1 shows the count of health incident events throughout Little Rock; Panel 2 shows the local Moran's $I$ statistics at each grid cell, the final panel shows areas that exhibit statistically significant clustering \citep{gimond2019}. 

\begin{figure}[!ht]
\centering
\includegraphics[width=0.75\linewidth]{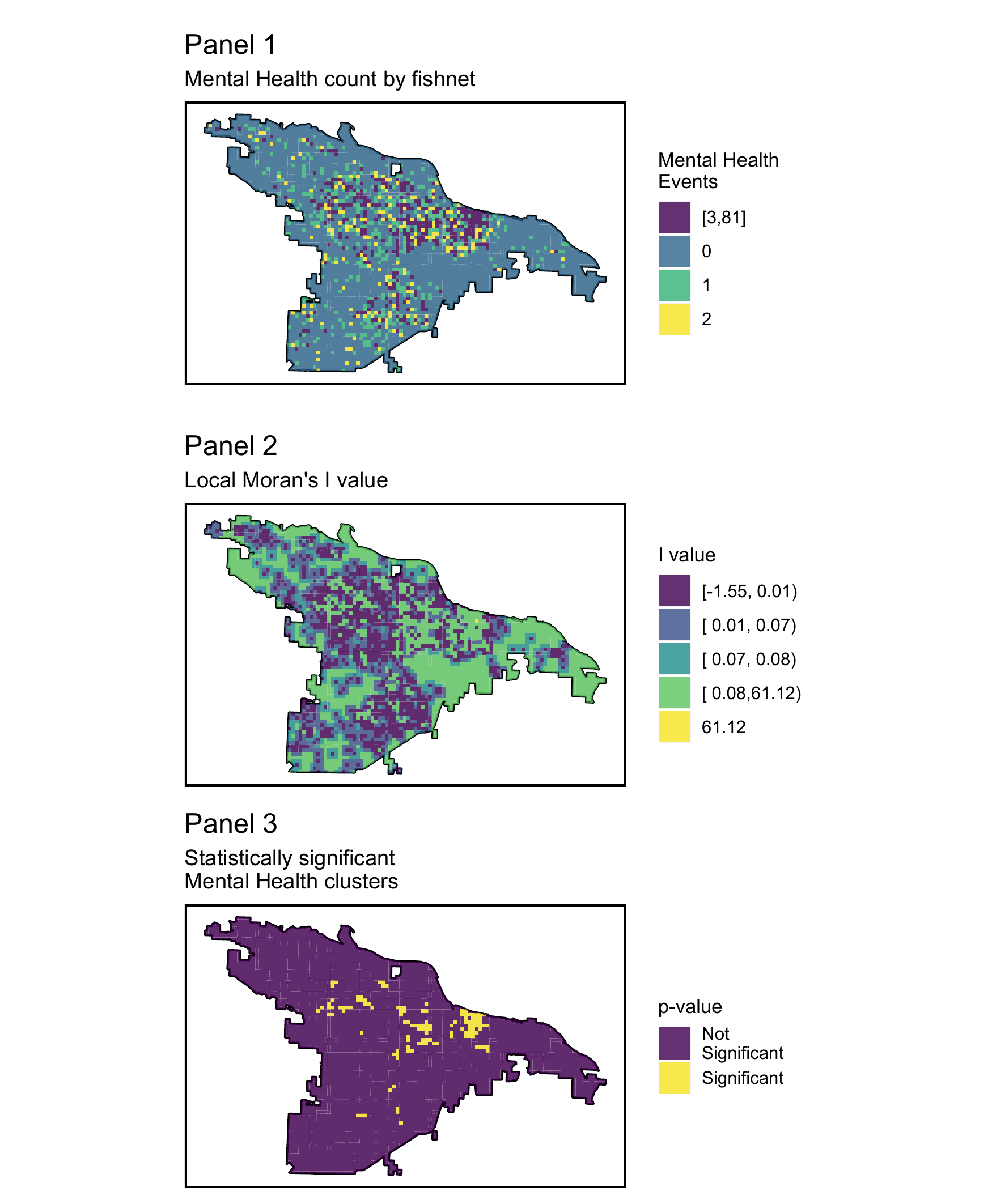}
\caption{Local Moran's I plot illustrating the spatial clusters of mental health incident calls in Little Rock, AR.}
\label{fig:local_morans}
\end{figure}

The presence of positive and significant spatial auto-correlation in the mental health incidents data clearly substantiates our claim that such events are clustered in space, instead of uniformly distributed over the entire region of interest. Having obtained such results is essentially the first step in the process of identifying a proper model \citep{Pebesma2019Spatial}. 

\subsubsection{Performance Comparison}

We compare the predictive performance of the four candidate methods in Table \ref{tab:modcomp}, and report the mean and standard deviation for each error measure. To better assess the accuracy of the models, we use four different error measures: Mean Absolute Percentage Error (MAPE), Mean Absolute Error (MAE) \& Root Mean Square Error (RMSE), see Table \ref{tab:modcomp}. The errors were calculated in a supervised learning set-up, where both Poisson regression and Random Forest models were built using leave-one-group-out cross-validation with the number of folds being equal to five. Below, we define the different error measures used to compare and describe the best performing model according to that criterion. 

\begin{table}[!ht]
\centering
\footnotesize{
\begin{tabular}{| c | llllll|}
\hline
               & \begin{tabular}[c]{@{}l@{}}MAPE\\  Mean\end{tabular} & \begin{tabular}[c]{@{}l@{}}MAPE \\ SD\end{tabular} & \begin{tabular}[c]{@{}l@{}}MAE \\ Mean\end{tabular} & \begin{tabular}[c]{@{}l@{}}MAE\\  SD\end{tabular} & \begin{tabular}[c]{@{}l@{}}RMSE \\ Mean\end{tabular} & \begin{tabular}[c]{@{}l@{}}RMSE\\ SD\end{tabular} \\
\hline
Poisson GLM   & 1.3112                                               & 0.0308                                             & 0.9098                                              & 0.2699                                            & 2.9166                                               & 1.5893                                            \\
\hline
Random Forest  & 1.306                                                & 0.0346                                             & 0.8677                                              & 0.1708                                            & 2.1904                                               & 0.9008                                            \\
\hline
Manski Model  & 1.302                                                 & NA                                             & 0.7708	                                              & NA                                            & 2.5832                                               & NA                                            \\
\hline
Spatial Durbin & 1.316                                                & NA                                                 & 0.6356                                              & NA                                                & 2.135                                                & NA    \\
\hline                                           
\end{tabular}}
\caption{Model performance comparison.}
\label{tab:modcomp}
\end{table}
First, the Mean Absolute percentage Error (MAPE) statistic captures the model's accuracy in terms of percentage error. The MAPE is calculated using the following formula:
\[
MAPE = \frac{1}{n} \sum^{n}_{i = 1} \abs{\frac{A_i - F_i}{A_i}} \times 100,
\]
where $A_i$ is the $i^{th}$ actual observation and $F_i$ is the $i^{th}$ forecast value. Since the MAPE expresses the error as percentage, it can be relatively easier to interpret when compared to other statistic measures.  The lower the percentage error, the more accurate the model represents the data. For a given model, it can be concluded that on average, the forecast is off by the MAPE. We can clearly see that on average all models forecasts were off by approximately 1.3\% with a standard deviation of approximately 0.0308 and 0.0346 for the Poisson GLM and Random Forest respectively. In terms of MAPE, all models perform relatively the same with the Manski model having the smallest MAPE. 

The Mean Absolute Error (MAE) statistic captures on average how large the forecast error is expected. The MAE is given by the formula
\[
MAE = \frac{ \sum^{n}_{i = 1} \abs{A_i - F_i} }{n},
\]
where $A_i$ is the $i^{th}$ actual observation and $F_i$ is the $i^{th}$ forecast value. Spatial Durbin error model had on average the smallest forecast error of 0.6356 followed by the Manski Model with a MAE of 0.7708  and Poisson GLM having the largest forecast error of 0.9098.

The Root Mean Square Error (RMSE) or otherwise also known as the Root Mean Square Deviation calculates the square root of the average of the square errors. The RMSE measures the spread of the prediction errors. The RMSE is given by the formula 
\[
RMSE = \sqrt{ \frac{\sum^n_{i = 1} (F_i - A_i)^2}{n}}.
\]
Spatial Durbin Error model had the smallest RMSE value of 2.135 followed by Random Forest with a RMSE of 2.1904 and the Poisson GLM having the largest RMSE of 2.9166.

\subsubsection{Goodness of fit metrics}

\begin{table}[!ht]
\centering
\begin{tabular}{|l|llll|}
\hline
               & \begin{tabular}[c]{@{}l@{}}$R^2$\\ Mean\end{tabular} & \begin{tabular}[c]{@{}l@{}}$R^2$\\ SD\end{tabular} & \begin{tabular}[c]{@{}l@{}}LogDev\\ Mean\end{tabular} & \begin{tabular}[c]{@{}l@{}}LogDev\\ Sd\end{tabular} \\ \hline
Poisson glm    & 0.3927                                            & 0.1517                                          & 0.6141                                                & 0.0509                                              \\ \hline
Random Forest  & 0.3822                                            & 0.0582                                          & 0.5844                                                & 0.0403                                              \\ \hline
Manski Model   & 0.4366                                            & NA                                              & 0.6124                                                & NA                                                  \\ \hline
Spatial Durbin & 0.4735                                            & NA                                              & 0.7102                                                & NA                                                  \\ \hline
\end{tabular}
\caption{Model goodness of fit comparison.}
\label{tab:Gofit}
\end{table}

In terms of Goodness of fit metrics, the R squared ($R^2$) values and logarithmic deviance score were used to evaluate the models. The most common measure is perhaps the $R^2$ that represents the percentage of variation explained by the model, 
\[
R^2 = 1- \frac{\sum_i (y_i - \hat{y}_i)^2}{\sum_i (y_i - \bar{y})^2},\; \hat{y}_i \doteq \text{ predicted value of } y_i, \; \bar{y} = \text{ grand mean},
\]
thus a larger $R^2$ is indicative of a better model fit. Note that the Adjusted $R^2$ value was not computed as it is rather difficult to compute for random forest models and, thus, difficult to use in goodness-of-fit comparison. The Logarithmic Deviance score is a measure of the deviance between the predicted and observed counts, via the log likelihood ratio. To measure this, we calculate the likelihood ratio of the observed value and the predicted value based on a Poisson distribution. The goodness of fit reported here is the negative log of the probability density so a lower value indicates a better predictive ability. As seen in table \ref{tab:Gofit}, the spatial Durbin error model obtained the largest R square value followed by the Manski model. Note that despite having obtained the largest R square value \emph{i.e.,} the best model in terms of R square goodness of fit metric, it obtained the largest logarithmic deviance score thus the worst model in logarithmic deviance score goodness of fit metric for the mental health data. In terms of the Logarithmic Deviance score goodness of fit metric, the random forest model obtained the smallest score. This suggests that the random forest model had the smallest deviance between predicted and observed count of mental health incidents \emph{i.e.,} the best model of such category.

\begin{figure}[!ht]
\centering
\includegraphics[scale = 0.6]{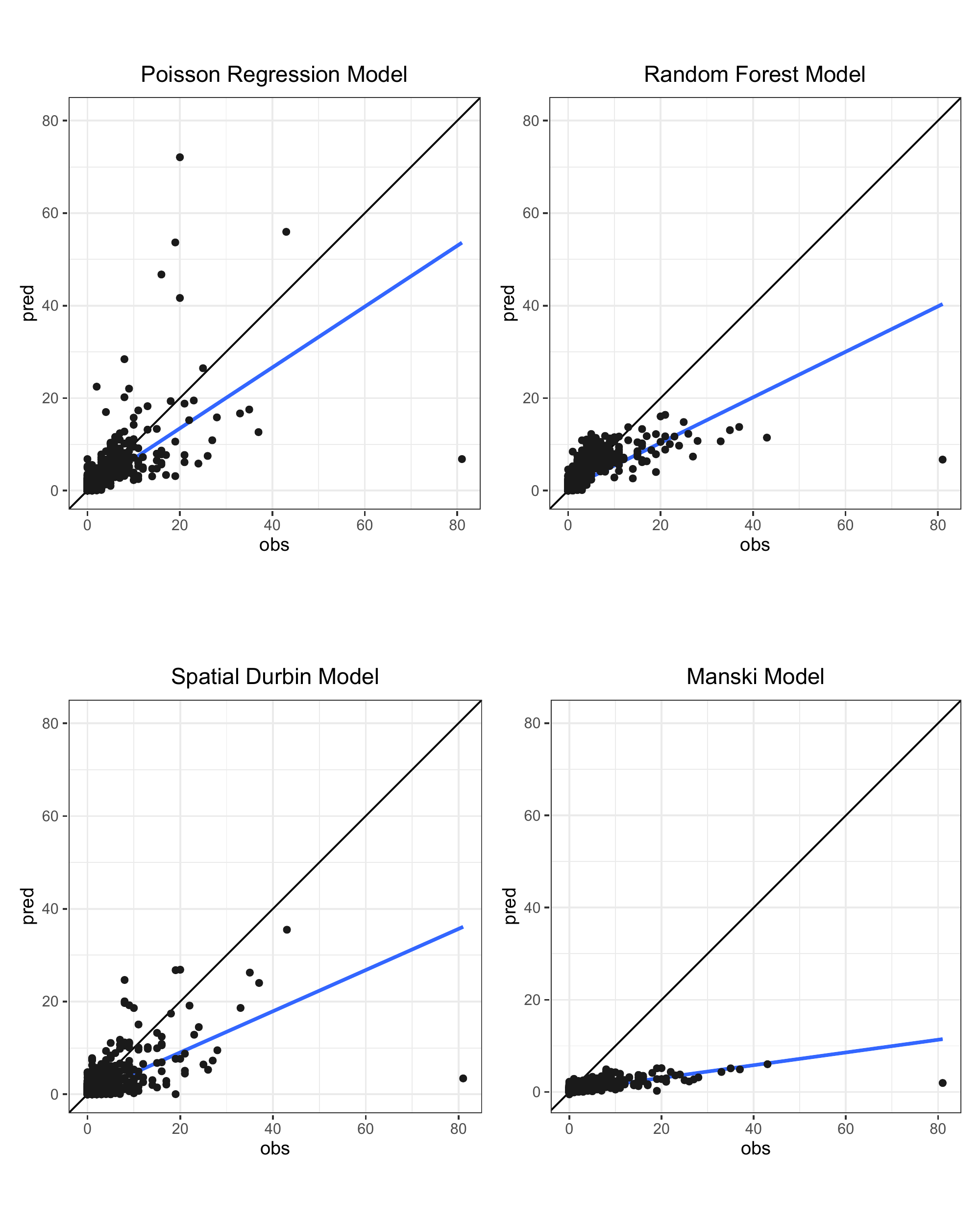}
\caption{Predicted versus observed mental health incident cases plots by the candidate models.}
\end{figure}


\subsubsection{Feature Importance Comparison}

Finally, we look at the important features or variables driving the prediction for each of the four candidate methods. We call these measures `variable importance' following the nomenclature used by random forest literature, but for purely statistical models such as Poisson regression or spatial Durbin models, the quantities being compared are a measure of each variable's significance. As discussed before, this a key step in the prediction process as the identification of important variables help us in determining which environmental and social features are predominantly occupying each of these predictive processes, investigate whether they play a risky or protective role and then allocate resources accordingly. 

A note about nomenclature for the features plotted on the following figures. There are three unique prefixes linked with each type of feature. Nearest neighbor (`NN') refers to features obtained by calculating the average distance between a fishnet grid cell centroid and its nearest neighbor in the Queen case definition. Euclidean distance (`ed') refers to features obtained by calculating the euclidean distance between a fishnet grid cell centroid and its first nearest neighbor. The prefix `agg' refers to the count of mental health incidents in a given fishnet grid cell. The term `agg' was coined based on the \emph{aggregate} function used in R to obtain the count of cases associated per fishnet cell. 

\begin{table}[!ht]
\centering
\scalebox{0.6}{
\begin{tabular}{|l|l|l|l|}
\hline
\rowcolor[HTML]{C0C0C0} 
Poisson\_GLM                         & Random\_Forest              
& Spatial\_Durbin                      & Manski                    
\\ \hline
agg\_Rentals\_Apts\_Over100units     & NN\_PoliceFacilities        
& agg\_Rentals\_Apts\_Over100units     & agg\_Rentals\_Apts\_Over100units
\\ \hline
agg\_Rentals\_Apts\_LessThan100units & NN\_Banks                   
& agg\_FastFoodAndBeverage             & agg\_FastFoodAndBeverage
\\ \hline
agg\_MajorDeptRetailDiscount         & agg\_BusStops               
& agg\_BusStops                        & agg\_BusStops
\\ \hline
agg\_FastFoodAndBeverage             & agg\_GasStationAndConvMart  
& agg\_Rentals\_Apts\_LessThan100units & agg\_GasStationAndConvMart
\\ \hline
agg\_MixedDrink\_BarRestClub         & agg\_FastFoodAndBeverage    
& agg\_GasStationAndConvMart           & agg\_MajorDeptRetailDiscount
\\ \hline
agg\_BusStops                        & NN\_ChildCareServices       
& agg\_MajorDeptRetailDiscount         & agg\_Rentals\_Apts\_LessThan100units
\\ \hline
NN\_ReligiousOrgs                    & NN\_BarberAndBeautyShops    
& agg\_HotelMotel.x                    & agg\_HotelMotel.x
\\ \hline
agg\_LiquorStores                    & NN\_ChildYouthServices      
& agg\_MixedDrink\_BarRestClub         & agg\_MixedDrink\_BarRestClub
\\ \hline
agg\_GasStationAndConvMart           & agg\_LiquorStores           
& NN\_Unsafe\_Vacant\_BldgsNEW         & NN\_Unsafe\_Vacant\_BldgsNEW
\\ \hline
NN\_Unsafe\_Vacant\_BldgsNEW         & NN\_ReligiousOrgs           
& agg\_LiquorStores                    & agg\_LiquorStores
\\ \hline
\end{tabular}
}
\caption{Top ten covariates with decreasing order of significance for each model.}
\label{tab: top10}
\end{table}

Table \ref{tab: top10} summarizes the top ten most influential features from each model. We note here that four similar features were found among the set of top features selected for each of the four models. These common features were: \texttt{agg\_FastFoodAndBeverage}, \texttt{agg\_BusStops}, \texttt{agg\_LiquorStores}, \texttt{agg\_GasStationAndConvMart}. As the four models highlight the importance of the influence these features had on the models, further interdisciplinary study involving experts from criminology and local law enforcement is required to understand whether any causal relationship exists between these environmental factors and mental health incidents in Little Rock, Arkansas.

\begin{figure}[!ht]
\centering
\includegraphics[height=4.5in]{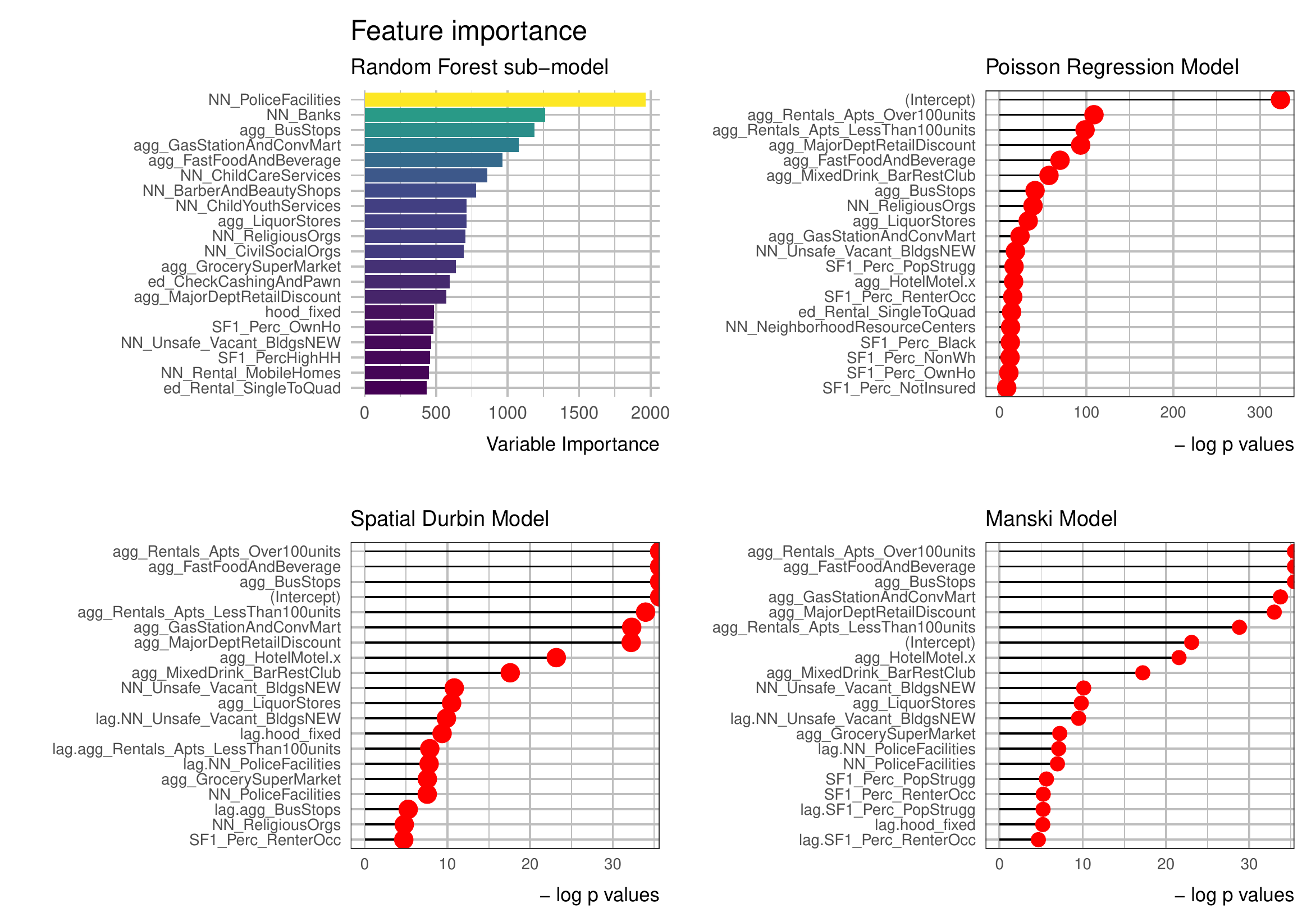}
\caption{Variable importance/significance for each model.}
\label{fig:var_imp}
\end{figure}

Figure \ref{fig:var_imp} illustrate the feature importance in descending order with respect to each model. In order to create a visual feature comparison between the random forest model feature importance and the remaining models, the $-\log_{10}P$-values of each predictor were plotted for the other three models. 

\section{Conclusion}\label{chap:end}

In this paper we used a machine learning framework to understand the effect of socio-demographics as well as environmental factors in predicting the spatial clusters of mental health incidents in Little Rock, Arkansas. The use of Moran's I for exploratory data analysis of existent spatial auto-correlation revealed an uneven distribution of mental health incidents across the area of study. The primary aim of this paper was to expand the methodology under RTM by incorporating statistical models to predict mental health incidents based on socio-economic predictors and environmental factors. We compared four different statistical methods' prediction accuracy and goodness of fit to provide insight into the list of factors affecting mental health incidents in Little Rock, Arkansas. Results indicate that in terms of prediction accuracy, the spatial econometric models (Manski and spatial Durbin error model) performed better than their model counterparts by a small margin. For model goodness of fit based upon R squared and Logarithmic Deviance score respectively, spatial Durbin error model and random forest model performed the best. The incorporation of these models under the risk terrain framework would definitely serve law enforcement agencies to properly allocate resources and address the unequal distribution of these mental health incidents. 

Furthermore, if law enforcement agencies adopt this framework, creating a meta model from the models generated may serve as a better tool if indecisive of which model to select based on prediction accuracy or goodness of fit. In addition to creating a meta model, the implementation of temporal features and regularization parameters would provide potentially better prediction and model goodness of fit results. The U.S. Federal Government has shown interest in crime prediction with the the National Institute of Justice holding a Real-Time Crime Forecasting Challenge in 2017. Beyond the above, it would also be meaningful to determine how these associations or patterns changed in relation to the ongoing Covid-19 pandemic, where mental and behavioral health services are needed even more and police are often the first responders to these types of calls.



\bibliographystyle{biometrika}

\bibliography{comp-ref-2}

\begin{thebibliography}{24}
\expandafter\ifx\csname natexlab\endcsname\relax\def\natexlab#1{#1}\fi

\bibitem[{Anselin(1995)}]{LISA}
\textsc{Anselin, L.} (1995).
\newblock Local indicators of spatial association—lisa.
\newblock \textit{Geographical Analysis} \textbf{27}, 93--115.

\bibitem[{Bivand \& Wong(2018)}]{SPDEP}
\textsc{Bivand, R.} \& \textsc{Wong, D. W.~S.} (2018).
\newblock Comparing implementations of global and local indicators of spatial
  association.
\newblock \textit{TEST} \textbf{27}, 716--748.

\bibitem[{Bivand et~al.(2008)Bivand, Pebesma, Gomez-Rubio \&
  Pebesma}]{bivand2008applied}
\textsc{Bivand, R.~S.}, \textsc{Pebesma, E.~J.}, \textsc{Gomez-Rubio, V.} \&
  \textsc{Pebesma, E.~J.} (2008).
\newblock \textit{Applied spatial data analysis with R}, vol. 747248717.
\newblock Springer.

\bibitem[{Boots(2001)}]{Moran_Range}
\textsc{Boots, B.} (2001).
\newblock Spatial pattern, analysis of.
\newblock In \textit{International Encyclopedia of the Social \& Behavioral
  Sciences}, N.~J. Smelser \& P.~B. Baltes, eds. Oxford: Pergamon, pp. 14818 --
  14822.

\bibitem[{Breiman(2001)}]{breiman2001random}
\textsc{Breiman, L.} (2001).
\newblock Random forests.
\newblock \textit{Machine learning} \textbf{45}, 5--32.

\bibitem[{Caplan et~al.(2015)Caplan, Kennedy, Barnum \& Piza}]{caplan2015risk}
\textsc{Caplan, J.~M.}, \textsc{Kennedy, L.~W.}, \textsc{Barnum, J.~D.} \&
  \textsc{Piza, E.~L.} (2015).
\newblock Risk terrain modeling for spatial risk assessment.
\newblock \textit{Cityscape} \textbf{17}, 7--16.

\bibitem[{Caplan et~al.(2011)Caplan, Kennedy \& Miller}]{caplan2011risk}
\textsc{Caplan, J.~M.}, \textsc{Kennedy, L.~W.} \& \textsc{Miller, J.} (2011).
\newblock Risk terrain modeling: Brokering criminological theory and gis
  methods for crime forecasting.
\newblock \textit{Justice Quarterly} \textbf{28}, 360--381.

\bibitem[{Chillar \& Drawve(2020)}]{chillar2020unpacking}
\textsc{Chillar, V.~F.} \& \textsc{Drawve, G.} (2020).
\newblock Unpacking spatio-temporal differences of risk for crime: An analysis
  in little rock, ar.
\newblock \textit{Policing: A Journal of Policy and Practice} \textbf{14},
  258--277.

\bibitem[{de~Jong et~al.(1984)de~Jong, Sprenger \& Veen}]{Extreme_Morans}
\textsc{de~Jong, P.}, \textsc{Sprenger, C.} \& \textsc{Veen, F.} (1984).
\newblock On extreme values of moran's i and geary's c ( spatial
  autocorrelation).
\newblock \textit{Geographical Analysis} \textbf{16}, 17--24.

\bibitem[{Drawve(2016)}]{drawve2016metric}
\textsc{Drawve, G.} (2016).
\newblock A metric comparison of predictive hot spot techniques and rtm.
\newblock \textit{Justice Quarterly} \textbf{33}, 369--397.

\bibitem[{Elhorst(2014)}]{Elhorst2014Spatial}
\textsc{Elhorst, J.} (2014).
\newblock \textit{Spatial econometrics: from cross-sectional data to spatial
  panels}.
\newblock Springer.

\bibitem[{Gimond(2019)}]{gimond2019}
\textsc{Gimond, M.} (2019).
\newblock {Intro to GIS and Spatial Analysis}.

\bibitem[{Gotway \& Stroup(1997)}]{gotway1997generalized}
\textsc{Gotway, C.~A.} \& \textsc{Stroup, W.~W.} (1997).
\newblock A generalized linear model approach to spatial data analysis and
  prediction.
\newblock \textit{Journal of Agricultural, Biological, and Environmental
  Statistics} , 157--178.

\bibitem[{James et~al.(2014)James, Witten, Hastie \&
  Tibshirani}]{James2014Introduction}
\textsc{James, G.}, \textsc{Witten, D.}, \textsc{Hastie, T.} \&
  \textsc{Tibshirani, R.} (2014).
\newblock \textit{An Introduction to Statistical Learning: With Applications in
  R}.
\newblock Springer Publishing Company, Incorporated.

\bibitem[{Lum \& Isaac(2016)}]{lum2016predict}
\textsc{Lum, K.} \& \textsc{Isaac, W.} (2016).
\newblock To predict and serve?
\newblock \textit{Significance} \textbf{13}, 14--19.

\bibitem[{Mohler et~al.(2013)}]{mohler2013modeling}
\textsc{Mohler, G.} et~al. (2013).
\newblock Modeling and estimation of multi-source clustering in crime and
  security data.
\newblock \textit{The Annals of Applied Statistics} \textbf{7}, 1525--1539.

\bibitem[{Mohler et~al.(2011)Mohler, Short, Brantingham, Schoenberg \&
  Tita}]{mohler2011self}
\textsc{Mohler, G.~O.}, \textsc{Short, M.~B.}, \textsc{Brantingham, P.~J.},
  \textsc{Schoenberg, F.~P.} \& \textsc{Tita, G.~E.} (2011).
\newblock Self-exciting point process modeling of crime.
\newblock \textit{Journal of the American Statistical Association}
  \textbf{106}, 100--108.

\bibitem[{Mohler et~al.(2015)Mohler, Short, Malinowski, Johnson, Tita, Bertozzi
  \& Brantingham}]{mohler2015randomized}
\textsc{Mohler, G.~O.}, \textsc{Short, M.~B.}, \textsc{Malinowski, S.},
  \textsc{Johnson, M.}, \textsc{Tita, G.~E.}, \textsc{Bertozzi, A.~L.} \&
  \textsc{Brantingham, P.~J.} (2015).
\newblock Randomized controlled field trials of predictive policing.
\newblock \textit{Journal of the American statistical association}
  \textbf{110}, 1399--1411.

\bibitem[{Montgomery et~al.(2006)Montgomery, Peck \& Vining}]{Montgomery}
\textsc{Montgomery, D.~C.}, \textsc{Peck, E.~A.} \& \textsc{Vining, G.~G.}
  (2006).
\newblock \textit{Introduction to Linear Regression Analysis (4th ed.)}.
\newblock Wiley \& Sons.

\bibitem[{Pebesma \& Bivand(2019)}]{Pebesma2019Spatial}
\textsc{Pebesma, E.~J.} \& \textsc{Bivand, R.} (2019).
\newblock {Spatial Data Science}.

\bibitem[{Perry et~al.(2013)Perry, McInnis, Price, Smith \&
  Hollywood}]{perry2013predictive}
\textsc{Perry, W.~L.}, \textsc{McInnis, B.}, \textsc{Price, C.~C.},
  \textsc{Smith, S.} \& \textsc{Hollywood, J.~S.} (2013).
\newblock Predictive policing: Forecasting crime for law enforcement .

\bibitem[{Townsley et~al.(2000)Townsley, Homel \&
  Chaseling}]{townsley2000repeat}
\textsc{Townsley, M.}, \textsc{Homel, R.} \& \textsc{Chaseling, J.} (2000).
\newblock Repeat burglary victimisation: Spatial and temporal patterns.
\newblock \textit{Australian \& New Zealand journal of criminology}
  \textbf{33}, 37--63.

\bibitem[{Wang \& Brown(2012)}]{wang2012spatio}
\textsc{Wang, X.} \& \textsc{Brown, D.~E.} (2012).
\newblock The spatio-temporal modeling for criminal incidents.
\newblock \textit{Security Informatics} \textbf{1}, 2.

\bibitem[{Wheeler \& Steenbeek(2021)}]{wheeler2021mapping}
\textsc{Wheeler, A.~P.} \& \textsc{Steenbeek, W.} (2021).
\newblock Mapping the risk terrain for crime using machine learning.
\newblock \textit{Journal of Quantitative Criminology} \textbf{37}, 445--480.

\end{thebibliography}

\end{document}